\documentclass[11pt,A4paper]{article}
\usepackage{subfigure}   
\usepackage{bbm}
\usepackage{tikz}
\usepackage{graphicx}
\makeindex

\begin{document}

\title{\bf Exceptional fuzzy spaces and octonions}

\author{Denjoe O'Connor\footnote {\tt denjoe@stp.dias.ie} and Brian P. Dolan\footnote{ {\tt bdolan@stp.dias.ie}}\\
\textit{School of Theoretical Physics }\\
\textit{10, Burlington Rd.}\\
\textit{Dublin 4}\\
\textit{Ireland}\\
}

\maketitle

\vspace{-12cm}
\rightline{DIAS-STP-22-16}
\vspace{12cm}

\begin{abstract}  
We construct the fuzzy spaces based on the three non-trivial co-adjoint
orbits of the exceptional simple Lie group, $G_2$.

\end{abstract}



  \rightline{DIAS-STP-22-16}  

  \vskip 30pt 
\noindent To be published in Particles, Fields and Topology: Celebrating A.P. Balachandran, a Festschrift volume for A.P. Balachandran. 

\section{Introduction}
Bal has had a long standing interest in co-adjoint orbits,
non-commutative geometry and quantum space-times. In this contribution
we pull from all these ingredients, spiced with a smidgen of octonions,
to make a contribution that hopefully Bal and others finds
interesting.

We consider the co-adjoint orbits of the exceptional group $G_2$ and
construct fuzzy versions of these spaces. In many respects the
orbits are very simple.  One has holonomy $U(1)\times U(1)$ while two others have $U(2)$ as
their holonomy group and, as Bal has emphasised to us, conformally
compactified Minkowski space-time is also $U(2)$.  These two are both ten dimensional orbits, one of which is
an $S^2$ bundle over the quaternionic projective plane $\mathbf{HP}^2$ and the other
is a $\mathbf{CP}^2$ bundle over $S^6$ which is diffeomorphic but not isospectral to the $SO(7)$ quadric
$\mathbf{Q}^5$.

\section{$G_2$}
The 14-dimensional exceptional group $G_2$ has rank-2. Its Dynkin diagram

\centerline{\begin{tikzpicture}
    \draw (-0.1,0.1) -- (1.1,0.1);
    \draw (0,0) -- (1,0);
    \draw (-0.1,-0.1) -- (1.1,-0.1);
    \draw (-0.1,0) circle (1mm);
    \fill[black!100!white] (1.1,0) circle (1mm);
\end{tikzpicture}}

\noindent is not symmetric and $G_2$ therefore has no complex representations.
The smallest non-trivial irrep. is 7 dimensional and
we shall use an explicit representation for the 14 generators\cite{CCVOS}, here labelled $T_1,\cdots,T_{14}$:
{\scriptsize \begin{eqnarray*}
T_1=\left(
\begin{array}{ccccccc}
 0 & 0 & 0 & 0 & 0 & 0 & 0 \\
 0 & 0 & 0 & 0 & 0 & 0 & 0 \\
 0 & 0 & 0 & 0 & 0 & 0 & 0 \\
 0 & 0 & 0 & 0 & 0 & 0 & -1 \\
 0 & 0 & 0 & 0 & 0 & -1 & 0 \\
 0 & 0 & 0 & 0 & 1 & 0 & 0 \\
 0 & 0 & 0 & 1 & 0 & 0 & 0 \\
\end{array}
               \right)\quad
&T_2=\left(
\begin{array}{ccccccc}
 0 & 0 & 0 & 0 & 0 & 0 & 0 \\
 0 & 0 & 0 & 0 & 0 & 0 & 0 \\
 0 & 0 & 0 & 0 & 0 & 0 & 0 \\
 0 & 0 & 0 & 0 & 0 & 1 & 0 \\
 0 & 0 & 0 & 0 & 0 & 0 & -1 \\
 0 & 0 & 0 & -1 & 0 & 0 & 0 \\
 0 & 0 & 0 & 0 & 1 & 0 & 0 \\
\end{array}
  \right)\\
  T_3=\left(
\begin{array}{ccccccc}
 0 & 0 & 0 & 0 & 0 & 0 & 0 \\
 0 & 0 & 0 & 0 & 0 & 0 & 0 \\
 0 & 0 & 0 & 0 & 0 & 0 & 0 \\
 0 & 0 & 0 & 0 & -1 & 0 & 0 \\
 0 & 0 & 0 & 1 & 0 & 0 & 0 \\
 0 & 0 & 0 & 0 & 0 & 0 & -1 \\
 0 & 0 & 0 & 0 & 0 & 1 & 0 \\
\end{array}
  \right) \quad
  &T_4=\left(
\begin{array}{ccccccc}
 0 & 0 & 0 & 0 & 0 & 0 & 0 \\
 0 & 0 & 0 & 0 & 0 & 0 & 1 \\
 0 & 0 & 0 & 0 & 0 & 1 & 0 \\
 0 & 0 & 0 & 0 & 0 & 0 & 0 \\
 0 & 0 & 0 & 0 & 0 & 0 & 0 \\
 0 & 0 & -1 & 0 & 0 & 0 & 0 \\
 0 & -1 & 0 & 0 & 0 & 0 & 0 \\
\end{array}
  \right)\\
  T_5=\left(
\begin{array}{ccccccc}
 0 & 0 & 0 & 0 & 0 & 0 & 0 \\
 0 & 0 & 0 & 0 & 0 & -1 & 0 \\
 0 & 0 & 0 & 0 & 0 & 0 & 1 \\
 0 & 0 & 0 & 0 & 0 & 0 & 0 \\
 0 & 0 & 0 & 0 & 0 & 0 & 0 \\
 0 & 1 & 0 & 0 & 0 & 0 & 0 \\
 0 & 0 & -1 & 0 & 0 & 0 & 0 \\
\end{array}
  \right)\quad
  &T_6=\left(
\begin{array}{ccccccc}
 0 & 0 & 0 & 0 & 0 & 0 & 0 \\
 0 & 0 & 0 & 0 & 1 & 0 & 0 \\
 0 & 0 & 0 & -1 & 0 & 0 & 0 \\
 0 & 0 & 1 & 0 & 0 & 0 & 0 \\
 0 & -1 & 0 & 0 & 0 & 0 & 0 \\
 0 & 0 & 0 & 0 & 0 & 0 & 0 \\
 0 & 0 & 0 & 0 & 0 & 0 & 0 \\
\end{array}
  \right)\\
  T_7=\left(
\begin{array}{ccccccc}
 0 & 0 & 0 & 0 & 0 & 0 & 0 \\
 0 & 0 & 0 & -1 & 0 & 0 & 0 \\
 0 & 0 & 0 & 0 & -1 & 0 & 0 \\
 0 & 1 & 0 & 0 & 0 & 0 & 0 \\
 0 & 0 & 1 & 0 & 0 & 0 & 0 \\
 0 & 0 & 0 & 0 & 0 & 0 & 0 \\
 0 & 0 & 0 & 0 & 0 & 0 & 0 \\
\end{array}
  \right)\quad
  &T_8=\frac{1}{\sqrt 3}\left(
\begin{array}{ccccccc}
 0 & 0 & 0 & 0 & 0 & 0 & 0 \\
 0 & 0 & -2 & 0 & 0 & 0 & 0 \\
 0 & 2 & 0 & 0 & 0 & 0 & 0 \\
 0 & 0 & 0 & 0 & 1 & 0 & 0 \\
 0 & 0 & 0 & -1 & 0 & 0 & 0 \\
 0 & 0 & 0 & 0 & 0 & 0 & -1 \\
 0 & 0 & 0 & 0 & 0 & 1 & 0 \\
\end{array}
  \right)\\
  T_9=\frac{1}{\sqrt 3}\left(
\begin{array}{ccccccc}
 0 & -2 & 0 & 0 & 0 & 0 & 0 \\
 2 & 0 & 0 & 0 & 0 & 0 & 0 \\
 0 & 0 & 0 & 0 & 0 & 0 & 0 \\
 0 & 0 & 0 & 0 & 0 & 0 & 1 \\
 0 & 0 & 0 & 0 & 0 & -1 & 0 \\
 0 & 0 & 0 & 0 & 1 & 0 & 0 \\
 0 & 0 & 0 & -1 & 0 & 0 & 0 \\
\end{array}
  \right)\quad
  &T_{10}=\frac{1}{\sqrt 3}\left(
\begin{array}{ccccccc}
 0 & 0 & -2 & 0 & 0 & 0 & 0 \\
 0 & 0 & 0 & 0 & 0 & 0 & 0 \\
 2 & 0 & 0 & 0 & 0 & 0 & 0 \\
 0 & 0 & 0 & 0 & 0 & -1 & 0 \\
 0 & 0 & 0 & 0 & 0 & 0 & -1 \\
 0 & 0 & 0 & 1 & 0 & 0 & 0 \\
 0 & 0 & 0 & 0 & 1 & 0 & 0 \\
\end{array}
  \right)\\
  T_{11}=\frac{1}{\sqrt 3}\left(
\begin{array}{ccccccc}
 0 & 0 & 0 & -2 & 0 & 0 & 0 \\
 0 & 0 & 0 & 0 & 0 & 0 & -1 \\
 0 & 0 & 0 & 0 & 0 & 1 & 0 \\
 2 & 0 & 0 & 0 & 0 & 0 & 0 \\
 0 & 0 & 0 & 0 & 0 & 0 & 0 \\
 0 & 0 & -1 & 0 & 0 & 0 & 0 \\
 0 & 1 & 0 & 0 & 0 & 0 & 0 \\
\end{array}
  \right)\quad
  &T_{12}=\frac{1}{\sqrt 3}\left(
\begin{array}{ccccccc}
 0 & 0 & 0 & 0 & -2 & 0 & 0 \\
 0 & 0 & 0 & 0 & 0 & 1 & 0 \\
 0 & 0 & 0 & 0 & 0 & 0 & 1 \\
 0 & 0 & 0 & 0 & 0 & 0 & 0 \\
 2 & 0 & 0 & 0 & 0 & 0 & 0 \\
 0 & -1 & 0 & 0 & 0 & 0 & 0 \\
 0 & 0 & -1 & 0 & 0 & 0 & 0 \\
\end{array}
  \right)\\
  T_{13}=\frac{1}{\sqrt 3}\left(
\begin{array}{ccccccc}
 0 & 0 & 0 & 0 & 0 & -2 & 0 \\
 0 & 0 & 0 & 0 & -1 & 0 & 0 \\
 0 & 0 & 0 & -1 & 0 & 0 & 0 \\
 0 & 0 & 1 & 0 & 0 & 0 & 0 \\
 0 & 1 & 0 & 0 & 0 & 0 & 0 \\
 2 & 0 & 0 & 0 & 0 & 0 & 0 \\
 0 & 0 & 0 & 0 & 0 & 0 & 0 \\
\end{array}
  \right)\quad
  &T_{14}=\frac{1}{\sqrt 3}\left(
\begin{array}{ccccccc}
 0 & 0 & 0 & 0 & 0 & 0 & -2 \\
 0 & 0 & 0 & 1 & 0 & 0 & 0 \\
 0 & 0 & 0 & 0 & -1 & 0 & 0 \\
 0 & -1 & 0 & 0 & 0 & 0 & 0 \\
 0 & 0 & 1 & 0 & 0 & 0 & 0 \\
 0 & 0 & 0 & 0 & 0 & 0 & 0 \\
 2 & 0 & 0 & 0 & 0 & 0 & 0 \\
\end{array}
\right).
\end{eqnarray*}}
The quadratic Casimir is $-\frac{1}{4}\sum_{a=1}^{ 14} T_a^2 = 2\, \mathbf {I}$, where
$\mathbf I$ is the $7 \times 7$ identity matrix.  Reference \cite{Macfarlane} gives many useful tensor identities for $G_2$.

For a general $7\times 7$ matrix we have the $G_2$ decomposition into irreps.
  \[ \mathbf 7 \times \mathbf 7 = \mathbf 1 + \mathbf 7 + \mathbf{14} +\mathbf{27},\]
  where the $\mathbf 7 + \mathbf{14}$ span all anti-symmetric $7\times 7$ matrices and $\mathbf{27}$ spans all traceless symmetric $7\times 7$ matrices.

  Irreps of $G_2$ are labelled by two Dynkin labels $[n_1,n_2]$ and the $\mathbf 7$ is $[1,0]$, the $\mathbf{14}$ is $[0,1]$ and the $\mathbf {27}$ is $[2,0]$.
In general $[n_1,n_2]$ has dimension

\smallskip

\noindent\scalebox{1.15}{${dim[n_1,n_2] = \frac{ (n_1+1)(n_2+1)(n_1+n_2+2)(n_1 + 2 n_2 +3)(n_1 + 3 n_2 + 4)(2 n_1 + 3 n_2 +5)}{5!}}$}

\medskip\noindent
and the quadratic Casimir is
\[C_2([n_1,n_2]) = \frac 1 3 n_1(n_1+5) +n_2(n_1 + n_2 +3).
\]

There are three different non-trivial co-adjoint orbits in the Lie algebra of $G_2$, \cite{Miyasaka}
\[ G_2/U(1)\times U(1), \quad G_2/U(2)_- \quad \mbox{and} \quad G_2/U(2)_+\, .\]
$U(2)_-$ and $U(2)_+$ here reflect two different ways of embedding
\[U(2) = [U(1)\times SU(2)]/{Z_2}\]
in $G_2$ (made explicit below) and here denoted\cite{BoyerAndGalicki,Miyaoka}
\begin{eqnarray}
  G_2/U(2)_-&=& G_2/ ([SU(2) \times U(1)]/Z_2)\\
  G_2/U(2)_+&=& G_2/([U(1) \times SU(2)]/Z_2).
\end{eqnarray}

As co-adjoint orbits these three spaces all lend themselves to a fuzzy construction.
$G_2/U(2)_-$ and $G_2/U(2)_+$ are both 10-dimensional manifolds and $G_2/U(2)_-$ is homeomorphic but not isometric to the complex quadric $\mathbf{Q}^5=SO(7)/SO(5) \times SO(2)$ (page 503 of Boyer et al\cite{BoyerAndGalicki}).
A fuzzy version of $\mathbf{Q}^5$ was previously
constructed in Dolan et al\cite{Q5}.
$G_2$, the automorphism group of the octonions, is a subgroup of $SO(7)$ and with the embedding $G_2 \hookrightarrow SO(7)$ the homogeneous space $SO(7)/G_2$ has $G_2$ holonomy and admits a metric with $SO(7)$ isometry, it is a squashed 7-sphere $SO(7)/G_2\approx S^7$. The tensor product of the 7-dimensional vector representation of $SO(7)$ with itself decomposes as
\[ \mathbf 7 \times \mathbf 7 = \mathbf 1 + \mathbf{21} + \mathbf{27}\]
where the $\mathbf {21}$ decomposes as $\mathbf 7 + \mathbf{14}$ under the embedding
$G \hookrightarrow SO(7)$.  
Denoting the 21 generators of $SO(7)$ in the 7-dimensional representation by
$X_1,\cdots,X_7,T_1,\cdots,T_{14}$, we can choose
{\scriptsize \begin{eqnarray*}
  X_1=\left(
           \begin{array}{ccccccc}
 0 & 0 & 0 & 0 & 0 & 0 & 0 \\
 0 & 0 & 1 & 0 & 0 & 0 & 0 \\
 0 & -1 & 0 & 0 & 0 & 0 & 0 \\
 0 & 0 & 0 & 0 & 1 & 0 & 0 \\
 0 & 0 & 0 & -1 & 0 & 0 & 0 \\
 0 & 0 & 0 & 0 & 0 & 0 & -1 \\
 0 & 0 & 0 & 0 & 0 & 1 & 0 \\
\end{array}
\right)\qquad &X_2=\left(
\begin{array}{ccccccc}
 0 & 0 & -1 & 0 & 0 & 0 & 0 \\
 0 & 0 & 0 & 0 & 0 & 0 & 0 \\
 1 & 0 & 0 & 0 & 0 & 0 & 0 \\
 0 & 0 & 0 & 0 & 0 & 1 & 0 \\
 0 & 0 & 0 & 0 & 0 & 0 & 1 \\
 0 & 0 & 0 & -1 & 0 & 0 & 0 \\
 0 & 0 & 0 & 0 & -1 & 0 & 0 \\
\end{array}
  \right)\\
   X_3=\left(
\begin{array}{ccccccc}
 0 & 1 & 0 & 0 & 0 & 0 & 0 \\
 -1 & 0 & 0 & 0 & 0 & 0 & 0 \\
 0 & 0 & 0 & 0 & 0 & 0 & 0 \\
 0 & 0 & 0 & 0 & 0 & 0 & 1 \\
 0 & 0 & 0 & 0 & 0 & -1 & 0 \\
 0 & 0 & 0 & 0 & 1 & 0 & 0 \\
 0 & 0 & 0 & -1 & 0 & 0 & 0 \\
\end{array}
               \right)\qquad
               &X_4=\left(
\begin{array}{ccccccc}
 0 & 0 & 0 & 0 & -1 & 0 & 0 \\
 0 & 0 & 0 & 0 & 0 & -1 & 0 \\
 0 & 0 & 0 & 0 & 0 & 0 & -1 \\
 0 & 0 & 0 & 0 & 0 & 0 & 0 \\
 1 & 0 & 0 & 0 & 0 & 0 & 0 \\
 0 & 1 & 0 & 0 & 0 & 0 & 0 \\
 0 & 0 & 1 & 0 & 0 & 0 & 0 \\
\end{array}
  \right)\\
   X_5=\left(
\begin{array}{ccccccc}
 0 & 0 & 0 & 1 & 0 & 0 & 0 \\
 0 & 0 & 0 & 0 & 0 & 0 & -1 \\
 0 & 0 & 0 & 0 & 0 & 1 & 0 \\
 -1 & 0 & 0 & 0 & 0 & 0 & 0 \\
 0 & 0 & 0 & 0 & 0 & 0 & 0 \\
 0 & 0 & -1 & 0 & 0 & 0 & 0 \\
 0 & 1 & 0 & 0 & 0 & 0 & 0 \\
\end{array}
\right)\qquad &X_6=\left(
\begin{array}{ccccccc}
 0 & 0 & 0 & 0 & 0 & 0 & 1 \\
 0 & 0 & 0 & 1 & 0 & 0 & 0 \\
 0 & 0 & 0 & 0 & -1 & 0 & 0 \\
 0 & -1 & 0 & 0 & 0 & 0 & 0 \\
 0 & 0 & 1 & 0 & 0 & 0 & 0 \\
 0 & 0 & 0 & 0 & 0 & 0 & 0 \\
 -1 & 0 & 0 & 0 & 0 & 0 & 0 \\
\end{array}
  \right)\\
X_7=\left(
\begin{array}{ccccccc}
 0 & 0 & 0 & 0 & 0 & -1 & 0 \\
 0 & 0 & 0 & 0 & 1 & 0 & 0 \\
 0 & 0 & 0 & 1 & 0 & 0 & 0 \\
 0 & 0 & -1 & 0 & 0 & 0 & 0 \\
 0 & -1 & 0 & 0 & 0 & 0 & 0 \\
 1 & 0 & 0 & 0 & 0 & 0 & 0 \\
 0 & 0 & 0 & 0 & 0 & 0 & 0 \\
\end{array}
\right). \hskip 12pt&
\end{eqnarray*}}
  These $X_i$ generators satisfy 
  \[ [X_i,X_j] = c_{i j}{}^a T_a + c_{i j}{}^k X_k.\]
 For  $i\ne j$ the anti-commutators satisfy
  \[ \{X_i,X_j\}=X_i X_j + X_j X_i= e_{i j} + e_{j i},\]
  where $e_{i j}$ is the matrix with $1$ in the $i-j$ position and zeros elsewhere and the anti-commutator is in the $\mathbf {27}$. We can define a projector that ${\mathcal P}$ that projects onto the  $\mathbf{7}$ of $SO(7)$ (or equivalently projecting out the $\mathbf{14}$ and the $\mathbf{27}$ of $G_2$),
  \[ {\mathcal P}([X_i,X_j])= c_{i j}{}^k X_k, \qquad  P(\{X_i,X_j\})= 0.\]
  For $i=j$
  \[X_i^2 = - \mathbf I + e_{i i}=-\frac 6 7 \mathbf I -
    \frac 1 7
    {\scriptsize
      \left(\begin{array}{ccccccc}
      1 & 0 & 0 & 0 & 0 &  0 & 0 \\
      0 & \ddots & 0 & 0 &\cdots & 0& 0 \\
      0 & \cdots &  1 & 0 & 0 &\cdots &  0\\
      0 & \cdots &  0 & -6 & 0 &\cdots &  0\\
      0 & \cdots & 0 & 0 & 1 & \cdots & 0\\
      0 & \cdots & 0 & 0 & 0 & \ddots & 0\\
      0 & 0 & 0 & 0 & 0 & 0 & 1\\
    \end{array}\right)}
    \qquad  (\mbox{no sum over}\ i)\]
  where $\mathbf I$ is the $7\times 7$ identity matrix
  and the $-6$ is in the $i-i$ position of the diagonal traceless matrix.
  Projecting out the $\mathbf {27}$ again
  \[{\mathcal P}(X_i^2) = -\frac 6 7 \mathbf I \]
  and
  \[ {\mathcal P}(X_i X_j)=-\frac 6 7 \delta_{i j} \mathbf I + \frac1 2 c_{i j}{}^k X_k.\]
  In fact the $c_{i j}{}^k$, which furnish a torsion tensor on the non-symmetric
  homogeneous space $SO(7)/G_2$, are the structure constants of the pure imaginary octonions
  and the non-associative algebra of octonions is obtained by matrix multiplication of the $X_i$ followed by projection onto the $\mathbf 1 + \mathbf 7$. The non-associativity of
  the octonion algebra in this construction arises from this projection. In our convention $c_{i j}{}^k=\epsilon_{ijk}$  is a completely anti-symmetric tensor with value 1 when $ijk = 123, 145, 176, 246, 257, 347, 365$ and the matrix $X_j$ has components $(X_i)_{j}{}^{k}=c_{ij}{}^k$.

\section{$G_2/U(2)_-$}

The generator $T_8$ commutes with $\{T_1,T_2,T_3;T_8\}$ which generate \hbox{$SU(2) \times U(1)$}
We denote the adjoint orbit of $T_8$, $g^{-1} T_8 g$ with $g\in G_2$, by $G_2/U(2)_-$, where $U(2)_- = [SU(2) \times U(1)]/Z_2$.  This adjoint orbit is related to the regular embedding of $SU(3)$ into $G_2$ where $SU(3)$ is generated by $\{T_1,\cdots,T_8\}$.
We can gain some insight into this structure by focusing on the point which is the origin of the orbit, $g = \mathbf 1$, which we shall call the N pole by analogy with $S^2 \approx SU(2)/U(1)$.  The 10 generators of $G_2$ that do not commute with $T_8$ are
$\{T_4, T_5, T_6, T_7,T_9,T_{10}, \cdots, T_{14}\}$
and these span the tangent space to $G_2/U(2)_-$ at the $N$ pole.
$G_2/SU(3)\approx S^6$ is a reduction of $G_2/U(2)_-$ under $U(2)_- \hookrightarrow SU(3)$ 
in the sense that $G_2/U(2)_-$ is a fibre bundle over $G_2/SU(3)$ (which is a squashed 6-sphere, $G_2/SU(3)\approx S^6$)  with a 4-dimensional fibre.  The adjoint action of $SU(3)$ on $T_8$ generates $\mathbf{CP}^2$, with
$\{T_4,T_5,T_6,T_7\}$ tangent to $\mathbf{CP}^2$ at the N pole, \cite{FuzzyCPN}
so the fibre is $\mathbf{CP}^2$. The remaining 6 generators $\{T_9,\cdots,T_{14}\}$ are tangent to the $S^6$ base at the N pole. In fact $G_2/U(2)_-$ is diffeomorphic, but not isometric, to the complex quadric $\mathbf {Q}^5$ (see Miyaoka \cite{Miyaoka}).
In summary we have
\[\begin{array}{ccc} \mathbf{CP}^2 & \longrightarrow & G_2/U(2)_-\approx \mathbf{Q}^5\\
    & & \\
                         & & \Big\downarrow \\
                                   & & \\& & G_2/SU(3)\approx{S^6},\end{array} \]
                               where again the equivalences on the total space and the base space are merely diffeomorphisms.

We can construct a rank 1 projector that commutes with the holonomy group $U(2)_-$ generated by $\{ T_1 , T_2, T_3; T_8\}$,
\[ P_1= \frac 1 2    {\scriptsize \left(
  \begin{array}{ccccccc}
    0 & 0 & 0 & 0 & 0 & 0 & 0 \\
    0 & 1 & -i & 0 & 0 & 0 & 0\\
    0 & i & 1 & 0 & 0 & 0 & 0 \\
    0 & 0 & 0 & 0  &0 & 0 & 0 \\
    0 & 0 & 0 & 0  &0 & 0 & 0 \\
    0 & 0 & 0 & 0  &0 & 0 & 0 \\
    0 & 0 & 0 & 0  &0 & 0 & 0
    \end{array}\right)}=\mathbf{I} +\frac{1}{2} \{X_2,X_3\}^2 +\frac{i}{6} X_1 -\frac{i}{2\sqrt{3}} T_8, \]
then $P_1$ commutes with $T_1,T_2,T_3,T_8$ but not with any linear
combination of the other $T_i$.  The metric and K\"ahler structure are
encoded into these projectors
\cite{FuzzyCPN,Dolan:2006tx,FuzzyGrassmannian}. Also the orbit of
$P_1$ under $SO(7)$ gives $\mathbf{Q}^5$ and at a generic point of
the orbit $P_1=\frac{1}{2}({\bf m}+i{\bf n})$ where ${\bf m}=-{\bf n}^2$
and ${\bf n}^3=-{\bf n}$.  Multinomials in the components of ${\bf n}$ are a basis for functions 
on $\mathbf{Q}^5$, in analogy with how the unit vector in $\mathbf{R}^3$ can be used to construct functions 
on $S^2$. The spectra of Laplacians on the orbits are given by the
quadratic Casimirs of the representations of $G_2$ (and $SO(7)$ for $\mathbf{Q}^5$) so the spectra are not the same: though ${\bf n}$ provides
the functions the manifolds are not isospectral.
 
As a co-adjoint orbit $G_2/U(2)_-$ is a symplectic space that admits a fuzzy description.
Under the chain of embeddings
\[ \begin{array}{ccccc}
 G_2  &\hookleftarrow& SU(3) &\hookleftarrow& U(2)_-\\
  \mathbf 7&\  \longrightarrow\ & \mathbf 1 + \mathbf 3 + \bar {\mathbf 3} &
                                                                           \ \longrightarrow\ & \mathbf 1_0 + (\mathbf {2}_1 + \mathbf {1}_{-2}) + (\mathbf 2_{-1} + \mathbf 1_{2})\\
     \mathbf {14} &\  \longrightarrow\ & \mathbf 3 + \bar{\mathbf 3} + \mathbf 8 &
     \ \longrightarrow\ & (\mathbf {2}_1 + \mathbf {1}_{-2}) + (\mathbf 2_{-1} + \mathbf 1_{2})
   + (\mathbf 1_0  + \mathbf 2_3 + \mathbf 2_{-3} + \mathbf 3_0).\end{array}
\] 
The $\mathbf 7$  contains a neutral singlet of $G_2/U(2)_-$, and so $\mathbf 7 \times \mathbf 7$ matrices provide a candidate for a matrix representation of functions on fuzzy  $G_2/U(2)_-$.  In terms of Dynkin labels the $\mathbf 7$ is $[1,0]$ and the symmetric product of $n$ of these is the $[n,0]$ with dimension
\[d_{n,0}=\frac{(n+1)(n+2)(n+3)(n+4)(2 n+5)}{5!}\, .\]
In analogy with the constructions in\cite{Q5,FuzzyCPN,Dolan:2006tx,FuzzyGrassmannian}, we propose $d_{n,0} \times d_{n,0}$ matrices as a fuzzy representation of functions on $G_2/U(2)_-$, with matrix multiplication giving a star product for multiplication of functions.

The Laplacian on a $d_{n,0}\times d_{n,0}$ matrix $\Phi$ representing a
function on fuzzy space $G_2/U(2)_-$ is obtained from $-\nabla^2 \Phi=\frac{1}{4} [T_a[T_a,\Phi]]$, with $T_a$ in the $d_{n,0}$ dimensional irrep. of $G_2$. Eigenvalues $\lambda$ of the Laplacian on $G_2/U(2)_-$ are therefore given by the second order Casimirs of the irreducible $G_2$ representations appearing in the product $[n,0]\times [n,0]$ representation. The first two of these, ordered with increasing $\lambda$, are
\[{\scriptsize
  \begin{array}{ccccccccccccccccccccc}
  [1,0]&\times &[1,0] &= &[0,0] &+&[1,0] &+& [0,1] &+& [2,0]& &  &&&&&&&& \\
  \mathbf{7}&\times& \mathbf{7}  & & \mathbf 1  &+&  \mathbf 7  &+ &\mathbf {14}  &+ &\mathbf {27}& &  &&&&&&&&\\
       &&\lambda:  &&   0 &&  2 &&   4 &&  \frac{14}{3} && \\
    \\
    \ [2,0]  &\times &[2,0]&= &[0,0] &+ &[1,0] &+& [0,1] &+ &2[2,0] &+&2[1,1]&+&[3,0]&&&&&&\\
    \mathbf {27} &\times &\mathbf {27} &=&  \mathbf 1   &+&  \mathbf 7   &+ & \mathbf {14} &+& 2(\mathbf {27})  &+&2(\mathbf{64})  &+&\mathbf{77} &&&&&&\\
       && \lambda:   & &   0 && 2 &&  4 && \frac{14}{3} && 7 &&8\\
    \\
&&&&&&&&   &+&\mathbf{77}'&+&\mathbf{189}&+&\mathbf{182}\\
&&&&&&&&       &+&[0,2]&+&[2,1]&+&[4,0]\\
&&&&&&&&    &&10&&\frac{32}{3}&&12.\\
  \end{array}}
\]
The bold numbers above the eigenvalues (the dimension of the relevant irrep.) are the degeneracies associated with the corresponding eigenvalue.
The Laplacian has a largest eigenvalue which grows as $n^2$, and the total number of eigenvalues is $d_{n,0}^2$, so for large $n$ \[ d_{n,0} \sim \frac{1}{60} n^5\] and reading the dimension of the manifold from Weyl's law \cite{Weyl:1911}
the space is of dimension 10, compatible with the fact
that $G_2/U(2)_-$ is a 5-dimensional complex manifold.

\section{$G_2/U(2)_+$}
The generator $T_3$ commutes with $\{T_3; T_8, T_9, T_{10}\}$, which generates
a $U(2)\subset G_2$ which we shall denote $U(2)_+$. The adjoint orbit of $T_3$, $g^{-1} T_3 g$ with $g\in G_2$, this is $G_2/U(2)_+$.
The 10 generators of $G_2$ that do not commute with $T_3$ are
$\{T_1, T_2, T_4, T_5, T_6, T_7, T_{11}, T_{12}, T_{13}, T_{14}\}$
and these span the tangent space to $G_2/U(2)_+$ at its $N$ pole.
There is a regular embedding of $SO(4) = [SU(2)\times SU(2)]_{Z_2}\hookrightarrow G_2$ and $G_2/SO(4)$ is an 8 dimensional manifold that is a reduction of $G_2/U(2)_+$ under $U(2)_+ \hookrightarrow SO(4)$ in the sense that $G_2/U(2)_+$ is a fibre bundle over $G/SO(4)$ with a 2-dimensional fibre.
At the N pole the generators of $SO(4)$ that are not in $U(2)_+$ are $\{T_1,T_2\}$ and these must be tangent to the fibre. The orbit of the adjoint action of the $SU(2)$ generated by $\{T_1,T_2,T_3\}$ acting on $T_3$ is a 2-sphere, so the fibre is a 2-sphere.
It is argued in \cite{Boya} that the 8 dimensional base $G_2/SO(4)$ is in fact diffeomorphic, but not isometric, to
the quaternionic projective plane $\mathbf{HP}^2 \approx Sp(3)/Sp(2)\times Sp(1)$, it admits a quaternionic K\"ahler structure \cite{Conti} and is an example of a Wolf space \cite{Wolf}.

So we have
\[\begin{array}{ccc} S^2\approx \mathbf{CP}^1 & \longrightarrow & G_2/U(2)_+\\
    & & \\
                         & & \Big\downarrow \\
                         & & \\& & G_2/SO(4)\approx \mathbf{HP}^2,\end{array} \]
                     with the equivalences on the total space and the base space being merely diffeomorphisms not isometries, the metrics are not equivalent.

As a co-adjoint orbit $G_2/U(2)_+$ is a symplectic space that admits a fuzzy description.
Under the chain of embeddings
\[ \begin{array}{ccccc}
 G_2  &\hookleftarrow& SO(4)\approx SU(2) \times SU(2) &\hookleftarrow& U(2)_+\\
  \mathbf 7&\  \longrightarrow\ &(\mathbf 1,\mathbf 3) + (\mathbf 2,\mathbf 2)&
                                                                                \ \longrightarrow\ &\mathbf {3}_0 + \mathbf 2_{1} + \mathbf 2_{-1}\\
     
    \mathbf {14}&\  \longrightarrow\ &(\mathbf 1,\mathbf 3) +(\mathbf 3,\mathbf 1) + (\mathbf 2,\mathbf 4)&
                                                                                \ \longrightarrow\ &\mathbf {3}_0 + \mathbf 1_{1,0} + \mathbf 1_{0,0}+ \mathbf 1_{-1,0}+\mathbf 4_{1} + \mathbf 4_{-1}\end{array}.\]
For a harmonic expansion of functions on $G_2/U(2)_+$ we need irreps of $G_2$ that contain a neutral singlet of the holonomy group $U(2)_+$, so a harmonic expansion should not contain a $\mathbf 7$ and the smallest non-trivial irrep that can appear in a harmonic expansion of a function on $G_2/U(2)_+$ is the $\mathbf {14}$.                                                                         
Tensor products of the three lowest dimensional non-trivial $G_2$ irreps. are                                       \begin{eqnarray*}
    \mathbf 7 \times \mathbf 7 &=& \mathbf 1 + \mathbf 7 +\mathbf {14} + \mathbf{27}\label{eq:7x7}\\
    \mathbf {14} \times \mathbf {14} &=& \mathbf 1 +\mathbf {14} + \mathbf{27}
     +\mathbf{77} + \mathbf{77}'\label{eq:14x14}\\
    \mathbf {27} \times \mathbf{27} &=&  \mathbf 1 + \mathbf 7 +\mathbf {14}
  + 2(\mathbf{27}) +2(\mathbf{64}) +\mathbf{77} + \mathbf{77}'+\mathbf{189} + \mathbf{182}  \label{eq:21x21} \end{eqnarray*}
and the lowest dimensional non-trivial matrix representations for a candidate to provide functions on fuzzy $G_2/U(2)_+$ is that of $\mathbf{14} \times \mathbf{14}$ matrices. Higher dimensional function expansions come from products of $\mathbf{14}$'s. In terms of Dynkin labels the $\mathbf {14}$ is $[0,1]$ and, taking $n$ of these, the $[0,n]$
has dimension
\[d_{0,n}=\frac{(n+1)(n+2)(2 n+3)(3 n+4)(3 n+5)}{5!}.\]
$d_{0,n} \times d_{0,n}$ matrix multiplication gives a star product for multiplication of functions on $G_2/U(2)_+$. For large $n$ the dimension scales as
\[ d_{0,n} \sim \frac{3}{20} n^5,\]
compatible with the fact that $G_2/U(2)_+$ is a 5-dimensional complex manifold\cite{Weyl:1911}.

We can construct a rank 2 projector that commutes with the holonomy group $U(2)_+$ generated by $\{ T_3 , T_8, T_9; T_{10}\}$,
\[ P_2= \frac 1 2    {\scriptsize \left(
  \begin{array}{ccccccc}
    0 & 0 & 0 & 0 & 0 & 0 & 0 \\
    0 & 0 & 0 & 0 & 0 & 0 & 0\\
    0 & 0 & 0 & 0 & 0 & 0 & 0 \\
    0 & 0 & 0 & 1  &-i & 0 & 0 \\
    0 & 0 & 0 & i  &1 & 0 & 0 \\
    0 & 0 & 0 & 0  &0 & 1 & -i \\
    0 & 0 & 0 & 0  &0 & i & 1
    \end{array}\right)}=\frac 1 2 \bigl(\{X_4,X_5\}^2 +  \{X_6,X_7\}^2\bigr) + i  T_3, \]
then $P_2$ commutes with $T_1,T_2,T_3,T_8$ but not with any linear combination of the other $T_i$.
On the $[0,2]$ representation we would use $P_2 \otimes_a P_2$ restricted to the $\mathbf{14}$, where $a$ denotes anti-symmetrisation.
Again the metric and K\"ahler structure are encoded into these projectors.
      
Eigenvalues $\lambda$ of the Laplacian on $G_2/U(2)_+$ are given by the second order Casimirs of the irreducible $G_2$ representations appearing in the product $[0,n]\times [0,n]$ representation. The first two of these, ordered with increasing $\lambda$, are
\[{\scriptsize\begin{array}{ccccccccccccccccccccccccccccccc}
  [0,1]&\times& [0,1] &=&
  [0,0] &+& [0,1] &+& [2,0] &+& [3,0] &+& [0,2] &&
  && && && &&\\
  \mathbf {14}  &\times& \mathbf {14} &=&
  \mathbf 1 &+&  \mathbf {14} &+& \mathbf{27} &+& \mathbf {77} &+& \mathbf {77}' &&
  && && && &&\\
  && \lambda:  &&
  0 &&  4 &&  \frac{14}{3} && 8 && 10 &&
  && && && &&\\
\\
  \  [0,2]  &\times& [0,2] &=&
     [0,0] &+& [0,1] &+& [2,0] &+& [3,0] &+& 2[0,2]\\
    \mathbf {77}' &\times& \mathbf {77}' &=&
  \mathbf{1} &+& \mathbf{14} &+& \mathbf{27} &+& \mathbf {77} &+& 2(\mathbf {77}')\\
 && \lambda:  &&
  0 &&  4&&  \frac{14}{3} && 8 && 10&\\
                \\
       &&     &&    &+&[2,1] &+& [4,0] &+& [5,0]      &+& 2[3,1]\\ 
&& &&   &+&  \mathbf{189} &+& \mathbf{182} &+& \mathbf{378}  &+& 2(\mathbf{448})\\
                &&            && &&  \frac{32}{3} && 12 && \frac{50}{3}&&  15\\
                \\
        &&&&&&          &+& [0,3] &+& [2,2] &+& [6,0] &+& [3,2] &+& [0,4]\\
   &&&&&&   &+& \mathbf{273} &+& \mathbf{729} &+& \mathbf{714} &+&  \mathbf{1547} &+& \mathbf{748}\\
&&&&&&&& 18 && \frac{56}{3} && 22 && 24 && 28. \\              \end{array}}
\]

On inspection of the decomposition of the $[n,0]\times [n,0]$ and that of $[0,n]\times [0,n]$ one sees that the latter is a subset of the former, implying that the two spaces are not globally equivalent. A theorem of Nakata (see Boyer et al \cite{BoyerAndGalicki} and Nakata \cite{Nakata:2018}) establishes that in fact $\pi_3(G_2/U(2)_-)=\boldmath{Z}_3$ while $\pi_3(G_3/U(2)_+)=0$ further emphasising the global inequivalence, though it is stated in \cite{Miyaoka} that they are both diffeomorphic to $\mathbf{Q}^5$.

\section{$G_2/U(1)\times U(1)$}

 The combination $T_3 + T_8$ only commutes with $T_3$ and $T_8$ and so the orbit $g^{-1}(T_3 + T_8)g$, is $G_2/U(1)\times U(1)$. At the N pole the 12 generators 
$\{T_1, T_2 , T_4, T_5, T_6, T_7, T_9, T_{10}, T_{11}, T_{12}, T_{13}, T_{14}\}$
are tangent to \hbox{$G_2/U(1)\times U(1)$}.
$G_2/U(1) \times  U(1)$ is an $S^2$ bundle over $\mathbf{Q}^5$, at the N pole the generators orthogonal to the base are $\{T_1, T_2, T_3,  T_8\}$ and $\{T_1, T_2\}$ span the fibre while $\{T_4, ..., T_7, T_9, ..., T_{14}\}$ 
span the base, which is $G_2/U(2)_-$.
As an adjoint orbit $G_2/U(1)\times U(1)$ can be embedded in $\mathbf {R}^{14}$, spanned by the $14$ generators. Reducing this to $S^{13}$ we have the bundle
structure \cite{Miyaoka}
\[\begin{array}{ccc}
    S^1 & \longrightarrow & S^{13} \\
    &&\\
                         & & \Big\downarrow \\
&&\\
    S^2\approx \mathbf{CP}^1 & \longrightarrow & G_2/U(1)\times U(1)\\
    & & \\
                         & & \Big\downarrow \\
        & & \\
    \mathbf{CP}^2    &\longrightarrow  & G_2/U(2)_-\approx{\mathbf {Q}^5}\\
    & & \\
                            & & \Big\downarrow \\
        &&\\
&&    G_2/SU(3) \approx S^6\, .
  \end{array} \]
Furthermore, since $T_3$ and $T_8$ are in $SU(3)$, the holonomy can be enlarged and   
$G_2/U(1)\times U(1)$ can also be viewed as a bundle over $G_2/SU(3) \approx S^6$ with fibre $SU(3)/U(1)\times U(1)$.

Under the chain of embeddings
\[ \begin{array}{ccccc}
     G_2  &\hookleftarrow& SU(3) &\hookleftarrow& U(1)\times U(1)\\
     \hline  &&&&\\
     \mathbf 7&\  \longrightarrow\ & \mathbf 1 + \mathbf 3 + \bar{\mathbf 3} &  \ \longrightarrow\ & (0,0) +(1,1)+ (1,-1)+(0,-2)\\
&&&&   + (-1,-1) + (-1,1) + (0,2)\\
     &&&& \\
      \mathbf {14}&\  \longrightarrow\ & \mathbf 3 + \bar{\mathbf 3} + \mathbf 8 &
 \ \longrightarrow\ &
   (1,1) + (1,-1) + (0,-2)  \\
      &&&&  \hskip 40pt +  (-1,-1) + (-1,1) + (2,0)\\
     &&&& \kern -80pt+2(0,0)+(0,2)+(0,-2)+(3,1)+(-3,-1)+(3,-1)+(-3,1), 
   \end{array}\]
 where $(p,q)$ represents the charges of $U(1)\times U(1)$.

 However we have already seen that symmetric products of $\mathbf 7$ gives \hbox{$d_{n,0}\times d_{n,0}$} matrices, describing a space with 5 complex dimensions.
 A more likely candidate is the $[1,1]$, which is $\mathbf{64}$ dimensional.
    Taking $n$ of these the $[n,n]$ dimensional irrep. of
    $G_2$ has dimension
    \[ d_{n,n}=(n+1)^6.\]
    This large $n$ behaviour is suggestive if 6 complex dimensions which is compatible with $G_2/U(1)\times U(1)$.  We therefore propose $d_{n,n} \times d_{n,n}$ matrices as the fuzzy representation of functions on $G_2/U(1)\times U(1)$.

    There is no projector on the $\mathbf{7}\times \mathbf{7}$ representation that commutes with $(T_3,T_8)$ only,
    we need to go to a higher dimensional irrep. to construct the projector for $G_2/U(1)\times U(1)$.
    The $\mathbf{64}=[1,1]$ is in the product of three $[1,0]$'s, two of which are anti-symmetrised, $\mathbf{7} \times \mathbf{14} = \mathbf{7}+ \mathbf{27}+\mathbf{64}$, so a candidate for the projector in this case
    is $P_1 \otimes (P_2 \otimes_a P_2)$ on the $\mathbf {64}$.

        Eigenvalues $\lambda$ of the Laplacian on $G_2/U(1) \times U(1)$ are given by the second order Casimirs of the irreducible $G_2$ representations appearing in the product $[n,n]\times [n,n]$ representation.
    The number of irreps in the tensor product rapidly become very large and we give only the $n=1$ case

    \[{\scriptsize \begin{array}{ccccccccccccccccccccccccccccccccc}
 [1,1]&\times& [1,1] &=& [0,0] &+& [1,0] &+& 2[0,1] &+& 2[2,0] &+& 2[1,1] &+& 3[3,0]\\
 \mathbf {64} &\times& \mathbf {64}  &=& \mathbf 1  &+&  \mathbf {7} &+&2(\mathbf{14}) &+& \mathbf {27}  &+& 2(\mathbf{64}) &+& 3(\mathbf{77})\\
 && \lambda:&  &  0 && 2 && 4  && \frac{14}{3} && 7 && 8 \\
 \\
&&&& &+& 2[0,2] &+&3[2,1] &+& 2[4,0] &+& [1,2] &+& 2[3,1]\\
&&&&   &+& 2(\mathbf{77}')  &+& 3(\mathbf{189})  &+& 2(\mathbf{182})  &+& \mathbf{286}  &+& 2(\mathbf{448})\\
                     &&&&&& 10  && \frac{32}{3} && 12 && 14 && 15\\
                     \\
&&&&&& &+& [5,0] &+& [0,3] &+& [2,2]\\
                     &&&&&& &+& \mathbf{378}  &+& \mathbf{273}  &+& \mathbf{729}\\
&&&&&& && \frac{50}{3} && 18 && \frac{56}{3}.\\
                   \end{array}}\]

\section{Conclusions}
    
Our three families of fuzzy spaces are specified by matrices of dimension
$d_{n,0}$ for $G_2/U(2)_-$, $d_{0,n}$ for $G_2/U(2)_+$ and $d_{n,n}$ for $G_2/U(1) \times U(1)$ where the associated Laplacians are specified by
$-\nabla^2 \Phi=\frac{1}{4} [T_a[T_a,\Phi]]$ and $T_a$ the generator of
$G_2$ in the associated representations. The equivalence of the function algebras of $G_2/U(2)_{-}$ and $\mathbf{Q}^5$ is manifest in our construction, see \cite{Q5}. To tease out any relation of these with $G_2/U(2)_+$ would take more work
and we have not pursued that here. Furthermore one should be able to
construct fuzzy Dirac operators and equivariant vector bundles over
these spaces along the lines of Dolan et al \cite{Dolan:2006tx,Dolan:2007uf}. 

{\bf Acknowledgements:} We would like to thank Charles Nash for several helpful discussions.\par
\noindent
{\it Comments of Denjoe O'Connor}~:
I am delighted to have the opportunity to contribute to this
celebration of Bal's 85th year. It has been a pleasure knowing him for
the past 35 years since when we first met in Syracuse. Though we have few
joint publications he has been a significant influence over the
years. I especially enjoyed our productive meetings in Cinvestav,
Mexico where the conditions and scientific atmosphere for such meeting was
superb. More recently Bal has been a lively and insightful participant in
the new era of Zoom and especially in the DIAS, School of Theoretical Physics
seminar series. I hope we have many future years of productive collaboration.

\medskip
\noindent
{\it Comments of Brian Dolan}~:
I first met Bal more than 30 years ago in 1991 when he visited Lochlainn O'Raifeartaigh's group in the Dublin Institute for Advances Studies (DIAS).
People often gathered in the kitchen in Burlington Road and the conversation would wander over various topics, but Bal was usually
very quiet unless the conversation was about either physics or politics. If we strayed from physics for too long he would go silent for a while and then announce ``We should discuss'', which meant go to a blackboard and discuss physics. And his call was always answered.

I subsequently met Bal on a number of occasions: in Dublin, in Cinvestav in Mexico City and in Syracuse, as well as at various conferences elsewhere. His focus on physics was always intense, but he is also a very warm and helpful person. On one occasion I was visiting Syracuse for a few days and stayed with Bal and Indra. It came out over breakfast one morning that I had no research grant and that I had paid for my flight to the US from Ireland myself.  Bal immediately arranged that I be reimbursed for the travel from his own grant.

\section{Bibliography}\label{ra_secbib}\index{bibliography}

 \end{document}